# Similarity theory based on the Dougherty-Ozmidov length scale

Andrey A. Grachev,[a,b] * Edgar L Andreas,[c] Christopher W. Fairall,[b] Peter S. Guest[d] and P. Ola G. Persson[a,b]

[a] *NOAA Earth System Research Laboratory, Boulder, Colorado, USA*

[b] *Cooperative Institute for Research in Environmental Sciences, University of Colorado, Boulder, Colorado, USA*

[c] *NorthWest Research Associates, Inc., Lebanon, NH, USA*

[d] *Naval Postgraduate School, Monterey, CA, USA*

* Correspondence to: A. A. Grachev, NOAA Earth System Research Laboratory, 325 Broadway, R/PSD3, Boulder CO 80305-3337, USA, E-mail: Andrey.Grachev@noaa.gov

# Abstract

Local similarity theory is suggested based on the Brunt-Väisälä frequency and the dissipation rate of turbulent kinetic energy instead the turbulent fluxes used in the traditional Monin-Obukhov similarity theory. Based on dimensional analysis (Pi theorem), it is shown that any properly scaled statistics of the small-scale turbulence are universal functions of a stability parameter defined as the ratio of a reference height $z$ and the Dougherty-Ozmidov length scale which in the limit of $z$-less stratification is linearly proportional to the Obukhov length scale. Measurements of atmospheric turbulence made at five levels on a 20-m tower over the Arctic pack ice during the Surface Heat Budget of the Arctic Ocean experiment (SHEBA) are used to examine the behaviour of different similarity functions in the stable boundary layer. It is found that in the framework of this approach the non-dimensional turbulent viscosity is equal to the gradient Richardson number whereas the non-dimensional turbulent thermal diffusivity is equal to the flux Richardson number. These results are a consequence of the approximate local balance between production of turbulence by the mean flow shear and viscous dissipation. The turbulence framework based on the Brunt-Väisälä frequency and the dissipation rate of turbulent kinetic energy may have practical advantages for estimating turbulence when the fluxes are not directly available.

1.    **Introduction**

Sixty years ago Monin and Obukhov (1954) suggested a similarity theory which is the commonly accepted approach to describe turbulence in the near-surface atmosphere. The basis of the Monin-Obukhov similarity theory (MOST) had been laid earlier by Obukhov's (1946) fundamental paper (e.g. see historical survey by Foken, 2006). Among other things Obukhov (1946) proposed a buoyancy length scale $L$ ("Obukhov length") which plays a central role in MOST.

Based on dimensional analysis (Pi theorem), MOST states that turbulent fluxes of momentum and heat (in general case buoyancy) are the primary governing (independent) variables that, along with the buoyancy parameter $\beta$ define other (dependent) variables (e.g. vertical gradients, variances etc.) in the atmospheric surface layer on the height $z$. Original MOST was based on the assumption that the turbulent fluxes are constant with height and equal to the surface values in the layer conventionally called a surface or constant-flux layer. This is "surface scaling". Subsequently, Nieuwstadt (1984) demonstrated that in the stable boundary layer (SBL) the assumption of height-independent fluxes is not necessary and Monin-Obukhov similarity can be redefined in terms of the local fluxes at height $z$ (i.e., $z$-dependent fluxes) rather than on the surface values. Nieuwstadt's is called "local scaling". In fact, Nieuwstadt deprived the turbulent fluxes of their "privileged role" and paved the way to construct a local similarity theory in the SBL based on governing variables other than the fluxes.

The Pi theorem used in MOST provides only a general methodology, and the choice of the primary governing variables is not unique. Presumably, Smeets et al. (2000) first discussed a similarity theory based on non-MOST governing parameters. In their paper, Smeets et al.

modified MOST by replacing the friction velocity with the standard deviation of the vertical wind speed component $\sigma_w$ to study the SBL over a glacier surface in a predominantly katabatic flow.

Sorbjan (2006) proposed alternative local scaling for the SBL based on $\sigma_w$, the Brunt-Väisälä frequency $N$, and $\beta$ (a buoyancy length scale defined as $\sigma_w/N$) and introduced the concept of gradient-based scaling. Subsequently, Sorbjan (2008, 2010) suggested three more gradient-based scaling systems. The gradient-based similarity approach removes turbulent fluxes as governing parameters and replaces them with vertical gradients of mean wind speed and potential temperature. As a result, the gradient Richardson number, $Ri$, appears as a stability parameter instead of the Monin-Obukhov stability parameter $z/L$. Sorbjan (2006, 2008, 2010) and Sorbjan and Grachev (2010) discussed different universal functions plotted versus $Ri$ based on field data. Obukhov length $L$, the gradient Richardson number $Ri$, the Brunt-Väisälä frequency $N$, and other variables mentioned here will be defined in the Sections 3 and 4 below.

In this paper, we further develop Sorbjan's (2006) ideas and suggest a similarity theory for the stably stratified boundary layer based on $N$ and the dissipation rate of turbulent kinetic energy $\varepsilon$ (cf. Sorbjan and Balsley 2008). A buoyancy length scale constructed from $N$ and $\varepsilon$ was originally suggested by Dougherty (1961) and Ozmidov (1965) and herein is referred to as the Dougherty-Ozmidov length scale. It is also known as Ozmidov length and is widely used in oceanography to describe small-scale turbulence. In contrast to the gradient-based scaling, we consider various similarity functions versus both the Richardson number and a stability parameter defined as the ratio of a reference height $z$ and the Dougherty-Ozmidov length scale, which plays the role of the Obukhov length in the proposed approach. We use the extensive measurements of atmospheric turbulence from the Surface Heat Budget of the Arctic Ocean experiment (SHEBA) described in Section 2 to examine the Dougherty-Ozmidov length scale and to derive similarity functions.

## 2. Data and data processing

Turbulent measurements made over the Arctic pack ice during the Surface Heat Budget of the Arctic Ocean experiment (SHEBA) took place in the Beaufort Gyre from 2 October 1997 to 11 October 1998. Andreas et al. (2006, 2010a, 2010b, 2013), Persson et al. (2002), Persson (2012) and Grachev et al. (2005, 2007a, 2008, 2013) describe the SHEBA site, various measurements over the Arctic sea ice, data processing, accuracy of measurements, instrument calibration etc. Here we provide some relevant information about the turbulent and profile measurements in the near-surface atmosphere during SHEBA.

Turbulent statistics (fluxes, variances, spectra, cospectra) and mean meteorological data were continuously measured on a 20-m main tower at five levels, hereafter levels 1–5, nominally $z_1 \approx 2.2$ m, $z_2 \approx 3.2$ m, $z_3 \approx 5.1$ m, $z_4 \approx 8.9$ m, and $z_5 \approx 18.2$ (but 14 m during most of the winter). Each level of the tower was instrumented with identical Applied Technologies, Inc. (ATI), three-axis sonic anemometer/thermometers (K-probe) that sampled at 10 Hz and a Väisälä HMP-235 temperature and relative humidity (T/RH) probes. An Ophir fast-response infrared hygrometer was mounted on a 3-m boom at an intermediate level (about 8 m) just below level 4. Although a sonic anemometer/thermometer measures the so-called 'sonic' temperature, which is close to the virtual temperature, the moisture correction in sonic temperature is usually small for Arctic conditions (see estimate in Grachev et al., 2005, p. 205).

The 'slow' T/RH probes provided air temperature and relative humidity measurements at five levels and were used to evaluate the vertical temperature and humidity gradients. The mean wind speed and wind direction were derived from the sonic anemometers in a streamline coordinate system whereby we performed two rotations on the sonic measurements that forced the mean lateral and vertical wind speed components to zero (Kaimal and Finnigan, 1994, Sect. 6.6). Vertical gradients of the mean wind speed, potential temperature, and specific humidity that appear here were obtained by fitting a second-order polynomial through the 1-hr averaged profiles followed by evaluating the derivative with respect to $z$ for levels 1–5 (Grachev et al., 2005, their Eq. 8).

Hourly averaged turbulent fluxes and variances at each level were derived through the frequency integration of the appropriate cospectra and spectra, which were normally computed from seven overlapping 13.65-min data blocks (corresponding to $2^{13}$ data points) and then averaged over an hour (see other details in Persson et al., 2002). One-hour averaging intervals are required to reduce excessive data scatter in the similarity relationships. To separate the contributions from mesoscale motions to the calculated eddy-correlation fluxes, we applied a low-frequency cut-off at 0.0061 Hz (the sixth spectral value or a period of about 3 min) on the cospectra as a lower limit of integration; the upper limit of integration is 5 Hz, the Nyquist frequency. The low-frequency cut-off for turbulent contributions is chosen to lie in the spectral gap between the small- and large-scale contributions to the total transport [see spectra and cospectra plots in Grachev et al. (2005, Figure 8), Grachev et al. (2008, Figure 3), and Grachev et al. (2013, Figures 1-4)].

Several data-quality indicators based on objective and subjective methods have been applied to the original flux data (e.g., Grachev et al. 2007a, p. 319). In particular, to avoid a possible flux loss caused by inadequate frequency response and sensor separations, we omitted data with a local wind speed less than 1 m s$^{-1}$. In addition, data with a temperature difference between the air (at median level) and the snow surface less than 0.5°C have also been omitted to avoid the large uncertainty in determining the sensible heat flux in near isothermal conditions.

However, despite the data-quality control (QC), there are almost always outliers that are noticeably inconsistent with the rest of the dataset, in particular, because they are affected by other phenomena that are not described by similarity theory. To remove spurious or near-zero data points, we further checked the data prior to evaluating similarity functions to remove indeterminate forms such as zero divided by zero. Following the QC recommendations by Klipp and Mahrt (2004) and Sanz Rodrigo and Anderson (2013), we set minimum thresholds for the kinematic momentum flux (0.0002 m$^2$ s$^{-2}$), temperature flux (0.0002 K m s$^{-1}$), standard deviation of each wind speed component (0.01 m s$^{-1}$), standard deviation of air temperature (0.01 K), vertical gradients of mean velocity (0.001 s$^{-1}$) and mean temperature (0.001 K m$^{-1}$), and dissipation rate of turbulent kinetic energy (0.0003 m$^2$ s$^{-3}$). Note that the thresholds for the fluxes and standard deviation are also required to avoid amplitude resolution problems (Vickers and Mahrt, 1997, their Figure 1b).

Resolution of turbulent fluctuations for ATI sonic anemometers as well as for most other sonic anemometers (e.g. Young 81000, Gill WindMaster) is $u'$, $v'$, $w' \sim 0.01$ m s$^{-1}$ for wind speed components and $\theta' \sim 0.01$ K for sonic temperature. Thus, minimum thresholds of the momentum and temperature fluxes can be estimated as $<u'w'> \sim 10^{-4}$ m$^2$ s$^{-2}$ and $<w'\theta'> \sim 10^{-4}$ K m s$^{-1}$ respectively. Although our QC thresholds for the fluxes and standard deviation are less rigorous than those used by Klipp and Mahrt (2004) and Sanz Rodrigo and Anderson (2013), we also imposed additional restrictions on the gradient and flux Richardson numbers (see below).

In the current study, the dissipation rate of turbulent kinetic energy $\varepsilon$ was estimated based on a common method for measuring $\varepsilon$ in a turbulent flow that assumes the existence of an inertial subrange associated with a Richardson-Kolmogorov cascade. Note, that the various estimates of $\varepsilon$ are valid only for a locally isotropic inertial subrange (see Gargett et al., 1984; Albertson et al., 1997; Chamecki and Dias, 2004; Lien and D'Asaro, 2006 for discussion). The one-dimensional wavenumber energy spectrum of the longitudinal velocity component in the inertial subrange has the form

$$F_u(k) = \alpha \varepsilon^{2/3} k^{-5/3} , \qquad (1)$$

where $k$ is the wavenumber and $\alpha$ is the Kolmogorov constant ($\alpha \approx 0.5$-$0.6$; e.g., Kaimal and Finnigan, 1994); a value $\alpha = 0.55$ is adopted for our study.

Spatial scales and the wavenumber spectrum in (1) should be converted, respectively, into frequency scales and frequency spectrum, which is traditionally what a sonic anemometer measures. By using Taylor's frozen turbulence hypothesis, $k = 2\pi f / U$ (where $f$ is frequency and $U$ is mean wind speed), the wavenumber spectrum in the inertial subrange (1) can be written in term of frequency as follows:

$$S_u(f) = \alpha (U/2\pi)^{2/3} \varepsilon^{2/3} f^{-5/3} . \qquad (2)$$

Frequency $S_u(f)$ and wavenumber $F_u(k)$ spectra are related to each other through $f S_u(f) = k F_u(k)$ (Kaimal and Finnigan, 1994). Local isotropy assumes that spectra of lateral and vertical velocity components are 4/3 of the longitudinal velocity; that is,

$$S_v(f) = S_w(f) = (4/3) S_u(f) . \qquad (3)$$

Based on (2) and (3), we derived the dissipation rate of turbulent kinetic energy $\varepsilon$ in this study separately from the spectra for each velocity component ($u'$, $v'$, and $w'$) in the frequency domain 0.49 – 0.74 Hz located within the inertial subrange. We then took the median of these three values as the representative dissipation rate. With this procedure, we avoided the influence of possible spectral spikes on the estimation of the dissipation rate (see Figures 1 and 3 in Grachev et al., 2013) and reduced sampling error. Because our estimates of $\varepsilon$ are based on Eqs. (1)-(3), data without the Richardson-Kolmogorov cascade should be filtered out. This study follows Grachev et al. (2013) and imposes restrictions on the gradient and flux Richardson numbers, $Ri$ and $Rf$, such that we excluded data points if both $Ri$ and $Rf$ exceed a critical value 0.2 (see also Eq. (20) below). Applying this prerequisite filters out data points for which the $-5/3$ power law generally fails (Grachev et al., 2013, Figures 7, 8).

We also tested an alternative method to filter cases when the $-5/3$ Kolmogorov power law fails. Instead of the restrictions on the gradient and flux Richardson numbers, we imposed the following two prerequisites on the data. First, the data points where the spectral slope in the inertial subrange deviated more than 10% of the theoretical -5/3 slope were excluded from the analysis (cf. Hartogensis and De Bruin, 2005, where ±20% was used). Second, to restrict the influence of outliers on the bin-averaging, we imposed a prerequisite proposed and discussed by Grachev et al. (2008, 2012). Although these prerequisites differ from the restrictions imposed on the gradient and flux Richardson numbers by Grachev et al. (2013) and in the current study, these two approaches are generally equivalent (see Grachev et al., 2013, Figures 7 and 8).

## 3.     MOST formalism

MOST assumes that the kinematic turbulent momentum flux (or magnitude of the wind stress), $-<u'w'> = \tau$, and turbulent temperature flux, $<w'\theta'> = -H$, along with the buoyancy parameter, $\beta = g/\theta$, are the primary influential variables (a.k.a. governing, scaling, repeating variables or parameters) that 'control' the vertical variation of mean flow and turbulence characteristics in the atmospheric surface layer at height $z$. Thus, MOST can be considered as flux-based scaling (e.g. Sorbjan, 2010) where the scaling parameters are:

$$\tau, \quad H, \quad \beta. \tag{4}$$

This is the prime similarity hypothesis of Monin and Obukhov (1954).

The flux-based scaling parameters (4) uniquely define a system of three scales that represent length, velocity, and temperature:

$$L = \frac{\tau^{3/2}}{\kappa \beta H}, \quad u_* = \sqrt{\tau}, \quad \theta_* = \frac{H}{\sqrt{\tau}}. \tag{5}$$

The length scale $L$ in (5) is known as Obukhov length scale (Obukhov, 1946) where, historically, the von Kármán constant $\kappa \approx 0.4$ is included in the definition of $L$ simply by convention. Here and above $u_*$ is the friction velocity, $\theta$ is mean potential temperature, $g$ is the acceleration due to gravity, $u$ and $w$ are the longitudinal and vertical velocity components, respectively, $[']$ denotes fluctuations about the mean value, and $<\ >$ is a time or space averaging operator. The sign convention for the temperature flux is $H > 0$ in the SBL.

For simplicity, we consider the case of dry air; otherwise, in the buoyancy term, $\beta <w'\theta'>$, $\theta$ should be replaced by the virtual potential temperature $\theta_v$. Note, that all variables in this paper are expressed in a streamline coordinate system; therefore, $\tau = \tau_x = -<u'w'>$ represents the longitudinal (or downstream) component of the wind stress, whereas, the lateral (or crosswind) stress component $\tau_y = -<v'w'> = 0$ ($v'$ is lateral velocity components).

Variables that are not listed in (4) among the scaling parameters are considered as dependent variables. Consider the wind shear $\partial U / \partial z$. According to the MOST, the relevant physical variables for $\partial U / \partial z$ in the stationary, homogeneous atmospheric boundary layer adjacent to a horizontal plane are

$$\partial U / \partial z, \quad \tau, \quad H, \quad \beta, \quad z . \tag{6}$$

These five variables (n = 5) involve three fundamental dimensions: length, time, and temperature (k = 3). According to Buckingham's Pi theorem (e.g., Monin and Yaglom, 1971; Stull, 1988; Sorbjan, 1989; Barenblatt, 1996; Foken, 2006; Kramm and Fritz, 2009), there are n – k = 2 independent dimensionless $\pi$ groups representing the problem in the general form

$$\pi = f(\pi_1) . \tag{7}$$

This statement is also known as the first Pi theorem. The second Pi theorem states that each $\pi$ group in (7) is a function of k = 3 governing or repeating variables plus one of the remaining variables (the number of repeating variables is equal to the number of fundamental dimensions). In our case, the repeating variables are defined by (4). Note that the Pi theorem provides only a general approach, and the choice of dimensionless $\pi$ groups is not unique.

Using the flux-based governing parameters (4), we can now specify the $\pi$ groups in (7). The first $\pi$ group is based on the governing parameters (4) and $z$ that lead to the Monin-Obukhov stability parameter (Monin and Obukhov, 1954); that is, $\pi_1 = \zeta$, where

$$\zeta \equiv \frac{z}{L} = -\frac{z \kappa g <w'\theta'>}{u_*^3 \theta} \tag{8}$$

defined as the ratio of $z$ and the Obukhov length scale, see (5). The next dimensionless group involves the governing parameters (4) and the vertical gradient of mean wind speed that produces $\pi = \dfrac{L}{u_*} \dfrac{\partial U}{\partial z}$.

Now the functional relationship (7) for the non-dimensional vertical gradient of mean wind speed may be expressed as

$$\frac{L}{u_*}\frac{\partial U}{\partial z} = \varphi'_m(\zeta) \ . \tag{9}$$

It is convenient to replace (9) by the alternative form (Sorbjan, 1989)

$$\frac{\kappa z}{u_*}\frac{\partial U}{\partial z} = \varphi_m(\zeta) \ , \tag{10}$$

where $\varphi_m = \kappa \zeta \varphi'_m$. The von Kármán constant $\kappa$ on the left-hand side of Eq. (10) is conventionally introduced solely as a matter of convenience such that $\varphi_m(0) = 1$ for neutral conditions ($\zeta \equiv 0$).

Similarly to (6), the relevant physical variables for potential temperature lapse rate, $\partial \theta / \partial z$, on the height $z$ are assumed to be

$$\partial \theta / \partial z, \ \ \tau, \ \ H, \ \ \beta, \ \ z \ . \tag{11}$$

Just as in the previous case, the five independent variables (11) have three fundamental dimensions (i.e. n = 5 and k = 3) that lead to (7) where $\pi_1 = \zeta$ and $\pi = \frac{L}{\theta_*}\frac{\partial \theta}{\partial z}$. Similarly to (9), the non-dimensional vertical gradient of the mean potential temperature can be expressed as $\frac{L}{\theta_*}\frac{\partial \theta}{\partial z} = \varphi'_h(\zeta)$ that is eventually equivalent to

$$\frac{\kappa z}{\theta_*}\frac{\partial \theta}{\partial z} = \varphi_h(\zeta) \ , \tag{12}$$

where $\varphi_h = \kappa \zeta \varphi'_h$. For neutral conditions $\varphi_h(0) = Pr_{t0}$, where $Pr_{t0} \approx 1$ is a constant referred to as the neutral value of the turbulent Prandtl number, defined shortly.

Generally, MOST predicts that any properly scaled statistics of the turbulence at reference height $z$ are universal functions of the stability parameter (8), $\zeta = z/L$. Specifically, the standard deviation of wind speed components $\sigma_\alpha$ and air temperature $\sigma_t$ are scaled as

$$\frac{\sigma_\alpha}{u_*} = \varphi_\alpha(\zeta), \qquad \frac{\sigma_t}{|\theta_*|} = \varphi_t(\zeta), \qquad (13)$$

where $\alpha$ (= $u$, $v$, and $w$) denotes the longitudinal, lateral, or vertical velocity component. In addition, the dissipation rate of turbulent kinetic energy $\varepsilon$ in the frameworks of MOST can be expressed as

$$\frac{\kappa z \varepsilon}{u_*^3} = \varphi_\varepsilon(\zeta) . \qquad (14)$$

Other widely used stability parameters, along with (8), are the gradient Richardson number, $Ri$, defined by

$$Ri = \left(\frac{g}{\theta}\right) \frac{\partial \theta / \partial z}{(\partial U / \partial z)^2} = \frac{\zeta \varphi_h}{\varphi_m^2} \qquad (15)$$

and the flux Richardson number, $Rf$ (also known as the mixing efficiency) defined by

$$Rf = -\left(\frac{g}{\theta}\right) \frac{<w'\theta'>}{u_*^2 (\partial U / \partial z)} = \frac{\zeta}{\varphi_m} , \qquad (16)$$

where both $Ri$ and $Rf$ are expressed in a streamline coordinate system. The ratio of $Ri$ to $Rf$ is the turbulent Prandtl number:

$$Pr_t = \frac{K_m}{K_h} = \frac{<u'w'>(\partial \theta / \partial z)}{<w'\theta'>(\partial U / \partial z)} = \frac{Ri}{Rf} = \frac{\varphi_h}{\varphi_m} , \qquad (17)$$

where $K_m = -\dfrac{<u'w'>}{\partial U / \partial z}$ and $K_h = -\dfrac{<w'\theta'>}{\partial \theta / \partial z}$ are the turbulent viscosity and the turbulent thermal diffusivity, respectively.

The exact forms of the universal functions (10) and (12) – (14) are not predicted by MOST and must be determined from measurements. However, MOST predict the asymptotic behaviour of these functions under very stable ($\zeta \gg 1$) and extremely unstable stratification (free convection, $\zeta \ll -1$). In the very stable regime, stratification inhibits vertical motions and the turbulence no longer communicates significantly with the surface (e.g., Monin and Yaglom, 1971); thus, $z$ ceases to be a scaling parameter. This is $z$-less scaling. In this case, MOST predicts that various dimensional quantities become independent of $z$ (Obukhov, 1946; Monin and Obukhov, 1954). Specifically, the non-dimensional functions $\varphi'_m$, $\varphi'_h$, $\varphi_\alpha$, and $\varphi_t$ [see Eqs. (9), (13)] cannot contain $z$ in the definition and, therefore, asymptotically approach constant values when $\zeta \gg 1$ (cf., Nieuwstadt, 1984). For the non-dimensional functions $\varphi_m$, $\varphi_h$, and $\varphi_\varepsilon$, the $z$-less concept requires that $z$ cancels in Eqs. (10), (12), and (14); linear relationships result. Thus, in the $z$-less limit,

$$\varphi_x(\zeta) = \beta_x \zeta \ , \quad \varphi_\alpha(\zeta) = \beta_\alpha \ , \quad \varphi_t(\zeta) = \beta_t \ , \tag{18}$$

where $\beta_x$ ($x = m, h$, and $\varepsilon$), $\beta_\alpha$, and $\beta_t$ are numerical coefficients coefficients (not to be confused with the buoyancy parameter).

A simple linear interpolation provides blending between neutral and the $z$-less limits (18) for the $\varphi_x(\zeta)$:

$$\varphi_x(\zeta) = \alpha_x + \beta_x \zeta \ , \tag{19}$$

where generally $\alpha_m = \alpha_\varepsilon = 1$ and $\alpha_h = Pr_{t0}$. The universal functions $\varphi_\alpha(\zeta)$ and $\varphi_t(\zeta)$, Eq. (13), in the MOST framework are considered to be constant for all $\zeta > 0$.

Although, since the landmark 1968 Kansas field experiment (Businger et al., 1971), Eq. (19) has fit the available experimental data well for $\zeta < 1$ and measurements suggest $\beta_m \approx \beta_h \approx$ 5 (Högström, 1988; Sorbjan, 1989; Garratt, 1992; Handorf et al., 1999; Foken, 2008; Wyngaard, 2010), $z$-less scaling (18) has been questioned for stronger stability, including the limit of very stable stratification. Several studies reported that the stability functions $\varphi_m$ and $\varphi_h$ increase more slowly with increasing stability than predicted by the linear Eq. (19). A detailed review of the different non-linear similarity functions $\varphi_m$ and $\varphi_h$ based on data collected in a variety of conditions can be found in Sharan and Kumar (2011).

Several studies attempted to remove the ambiguity between predicted (z-less) and observed behaviour of the universal functions. Grachev et al. (2013) argued that the applicability of MOST (in the local scaling formulation) in stable conditions is limited by the inequalities

$$Ri < Ri_{cr} \quad \text{and} \quad Rf < Rf_{cr}, \tag{20}$$

where both critical values $Ri_{cr}$ and $Rf_{cr}$ are about 0.20-0.25. Various plots of the universal functions in the literature often contain data points which do not satisfy the condition (20); that is, they do not belong to MOST.

To evaluate different MOST functions, Grachev et al. (2013) suggested separating data points into subcritical and supercritical cases, that is, "separating the apples from the oranges" based on the prerequisite (20). According to Grachev et al. (2013, Figures 7 and 8), the upper limit of MOST in the SBL (20) coincides with the region for which the –5/3 Kolmogorov power law is applicable. In other words, the condition (20) also separates Kolmogorov and non-Kolmogorov turbulence in stratified turbulent shear flows. As mentioned in Section 2, in this study, both inequalities (20) with $Ri_{cr} = Rf_{cr} = 0.2$ have been imposed on the data to filter out cases when the Richardson-Kolmogorov cascade is not observed. This practice of separating the data into subcritical and supercritical regimes is consistent with laboratory experiments (e.g., Rohr et al., 1988), theoretical results (Baumert and Peters, 2004; Katul et al., 2014), and field measurements (e.g., Tjernström, 1993).

Figure 1 shows plots of the non-dimensional universal functions $\varphi_m$, Eq. (10), $\varphi_h$, Eq. (12), and $\varphi_\varepsilon$, Eq. (14) versus the Monin-Obukhov stability parameter for local scaling $\zeta = z/L$,

Eq. (8), when both prerequisites (20) with $Ri_{cr} = Rf_{cr} = 0.2$ have been imposed on the data. The individual 1-hr-averaged data shown in Figure 1 as the background x-symbols are a sample of the available data at one level and show the typical scatter of the data. According to the SHEBA data, numerical coefficients in (19) are $\beta_m = 5.0$ (Figure 1*a*), $\beta_h = 4.5$ (Figure 1*b*), and $\beta_\varepsilon = 5.0$ (Figure 1*c*). The numerical coefficients $\beta_m$, $\beta_h$, and $\beta_\varepsilon$ reported here are in close agreement with previously published results (see reviews by Yaglom, 1977; Högström, 1988; Sorbjan, 1989; Garratt, 1992; Hartogensis and De Bruin, 2005; Foken, 2008). For example, Kaimal and Finnigan (1994) recommend $\varphi_m = \varphi_h = \varphi_\varepsilon = 1 + 5\zeta$ for $0 < \zeta < 1$.

Grachev et al. (2013, their Figure 14) found a numerical coefficient of $\beta_w = 1.3$ in Eq. (13) for SHEBA. The neutral value of the turbulent Prandtl number in Eq. (19) for $x = h$ has not been specifically determined; instead, we accepted $Pr_{t0} = Pr_t = \beta_h / \beta_m = 0.9$ that coincides with our previous estimate $Pr_t = 0.9$ derived from a plot of $Pr_t$ versus $Ri$ in Sorbjan and Grachev (2010, their Figure 2). Further discussion on the turbulent Prandtl number in the SBL can be found in Grachev et al. (2007b), Anderson (2009), and references therein.

The universal functions $\varphi_m$ and $\varphi_\varepsilon$ discussed here are directly associated with the turbulente kinetic energy (TKE) equation (e.g. Garratt 1992; Kaimal and Finnigan 1994):

$$\partial <e> / \partial t = - <u'w'>(\partial U / \partial z) + \beta <w'\theta'> - \partial(<w'e> + <w'p'>/\rho)/\partial z - \varepsilon , \quad (21)$$

where $e = (u'^2 + v'^2 + w'^2)/2$ is TKE and $p'$ is the fluctuation in atmospheric pressure. Assuming steady state ($\partial <e> / \partial t = 0$), Eq. (21) reduces to

$$\varphi_m - \zeta - \varphi_T - \varphi_\varepsilon = 0 , \quad (22)$$

where $\varphi_T = (\kappa z / u_*^3) \partial (<w'e> + <w'p'>/\rho)/\partial z$ is the normalized vertical transport term, and other terms in (22) are defined by Eqs. (10), (8), and (14). The transport term $\varphi_T$ is generally neglected (e.g. Monin and Yaglom, 1971) and Eq. (22) can be written as

$$\varphi_m(1 - Rf) - \varphi_\varepsilon = 0, \qquad (23)$$

where $Rf$ is defined by (16). Note that subsequently specifying $\varphi_\varepsilon$ in Eq. (23), or in its modifications, leads to the, so-called, KEYPS or the O'KEYPS (Businger and Yaglom, 1971) equation for $\varphi_m$ (named after Obukhov, Kazansky, Ellison, Yamamoto, Panofsky, and Sellers) (Monin and Yaglom, 1971; Kramm et al., 1996; Katul et al., 2011).

In the 1968 Kansas data, Wyngaard and Coté (1971) found that under stable conditions, shear production and viscous dissipation are the dominant terms and they are essentially in balance, that is, $-<u'w'>(\partial U/\partial z) = \varepsilon$ or

$$\varphi_m = \varphi_\varepsilon . \qquad (24)$$

Equation (24), also means that the turbulent transport and the buoyancy production terms are either small or are generally in balance, $\varphi_T = \zeta$ (cf. Eq. (22)). Note, that the result $\varphi_\varepsilon \cong \varphi_m$ (or $\beta_\varepsilon \cong \beta_m$) has been known for a long time (at least since the landmark 1968 Kansas field experiment) and our data presented in Figure 1 agree with (24). In particular, Tjernström (1993, Figure 2) found that the balance $-<u'w'>(\partial U/\partial z) \approx \varepsilon$ is maintained up to $Ri < Ri_{cr} = 0.25$; however, the balance, Eq. (24), abruptly fails when $Ri$ exceed 0.25. In fact, according to our measurements, an actual difference between $\beta_m$ and $\beta_\varepsilon$ is within the accuracy of the experimental data. The result (24) will be used in Sec. 4.2 below.

Concluding this section, we note also that a more general formulation of similarity theory may include additional possible influences on the flux-gradient (or flux-variance, etc.)

relationships than the parameters listed in (4): for example, the Coriolis parameter, boundary layer depth, aerodynamic and scalar roughness lengths, molecular viscosity, and thermal conductivity. Such extra parameters would eventually lead to more $\pi$ groups in (7) (e.g., Barenblatt, 1996; Mahrt et al., 2003). Furthermore, Klipp and Mahrt (2004, Section 8*a*) formulated a generalized *z*-less similarity theory that contains the classical Monin-Obukhov *z*-less asymptote (18) as a special case. According to Klipp and Mahrt (2004), an additional variable $d\theta/dz$ should be added to the list (6) [or, equivalently, $\partial U/\partial z$ to the list (11)] but $z$ should be dropped to describe flux-profile relationships when $\zeta \gg 1$. Evidently, that in the frameworks of the Klipp-Mahrt approach, one may suggest that the variables (6) with an additional parameter $\partial \theta/\partial z$ will lead to (7) with an additional $\pi$ group on the right-hand side of the equation, $\pi_2 = \varphi'_h$ (n = 6 and k = 3). None of these cases are considered here.

## 4. The *N-ε* scaling

The flux-based scaling system (4) is not a unique combination of the governing parameters for describing stratified turbulent shear flows (Sorbjan, 2006, 2008, 2010). Here we derive universal functions based on a scaling system that includes the buoyancy frequency *N* (defined shortly) and the dissipation rate of turbulent kinetic energy $\varepsilon$ ("*N-ε* scaling").

### 4.1. *Dimensional analysis*

In oceanography, variables $\partial \theta/\partial z$, $\varepsilon$, and $\beta$ or, equivalently,

$$N, \quad \varepsilon, \quad \beta \tag{25}$$

are traditionally used as the governing parameters to describe small-scale turbulence. Here $N = \sqrt{\beta(\partial\theta/\partial z)}$ is the Brunt-Väisälä frequency, or buoyancy frequency. In the case of humid air or salt water, the buoyancy term $g(\partial\theta/\partial z)/\theta$ appearing in $N$ should be replaced by $g(\partial\theta_v/\partial z)/\theta_v$ or $-g(\partial\rho/\partial z)/\rho$ as discussed earlier ($\rho$ is the potential density). The oceanography community's use of the parameters (25) is primarily associated with the fact that they can be routinely measured in the ocean.

Similarly to (4) and (5), the scaling parameters (25) uniquely define a system of three fundamental scales for the length, velocity, and temperature:

$$L_{N\varepsilon} = \sqrt{\varepsilon/N^3}, \quad U_{N\varepsilon} = \sqrt{\varepsilon/N}, \quad \theta_{N\varepsilon} = \sqrt{\varepsilon N}/\beta. \tag{26}$$

Obviously $U_{N\varepsilon} = L_{N\varepsilon} N$ and $\theta_{N\varepsilon} = L_{N\varepsilon}(\partial\theta/\partial z)$. The buoyancy length scale $L_{N\varepsilon}$ in (26) was originally suggested by Dougherty (1961) and independently by Ozmidov (1965). Ironically, the length scale $L_{N\varepsilon}$ is widely known as the Ozmidov length scale (e.g. Dillon, 1982; Hunt et al., 1985; Rohr et al., 1988; Galperin et al., 1989; Baumert and Peters, 2000, 2004; Smyth and Moum, 2000; Sorbjan and Balsley, 2008; Mater et al., 2013). Historically the term "Ozmidov length scale" was introduced by Carl H. Gibson in the oceanographic community (R.V. Ozmidov personal communication, 1985). The buoyancy velocity and temperature (or density) scales (26) in the SBL were discussed by Gargett et al. (1984), Lee (1996), and Sorbjan and Balsley (2008).

Dougherty (1961) considered anisotropy of atmospheric turbulence at heights near 90 km and studied the ratio of $L_{N\varepsilon}$ to the Kolmogorov length scale (see also discussion by Lumley, 1964). Ozmidov (1965, his Eq. 5) constructed a buoyancy length scale from $\varepsilon$, $\partial\rho/\partial z$, $g/\rho$ to estimate vertical diffusivity in the ocean, and his formulation differs from the canonical relationship $L_{N\varepsilon} = \sqrt{\varepsilon/N^3}$ by only a numerical coefficient. The Dougherty-Ozmidov length scale $L_{N\varepsilon}$ is considered to define the size of the largest eddy that is unaffected by buoyancy in stratified turbulence (e.g., Gibson 1980).

Variables $\partial U / \partial z$, $\tau, H, \sigma_\alpha, \sigma_t$ etc., which are not listed in (26) among the scaling parameters, are considered as dependent variables in the framework of $N$-$\varepsilon$ scaling. Suppose we are interested in $\partial U / \partial z$ at height $z$; the relevant physical variables in this case are:

$$\partial U / \partial z, \quad N, \quad \varepsilon, \quad \beta, \quad z . \tag{27}$$

As in the case of the traditional MOST, with n = 5 and k = 3 [five independent variables (27) involving three fundamental dimensions], (7) leads to with $\pi_1 = z / L_{N\varepsilon} \equiv \xi$ and $\pi = \dfrac{L_{N\varepsilon}}{U_{N\varepsilon}} \dfrac{\partial U}{\partial z} =$

$\dfrac{1}{N} \dfrac{\partial U}{\partial z} = Ri^{-1/2}$, where the gradient Richardson number, $Ri$, is defined by (15). It is convenient to write a non-dimensional relationship for $dU/dz$ in the form

$$Ri = \psi_R(\xi) . \tag{28}$$

Thus the gradient Richardson number (15) is a universal function of a stability parameter $\xi = z / L_{N\varepsilon}$ defined as the ratio of a reference height $z$ and the Dougherty-Ozmidov length scale.

Applying the above formalism to the turbulent fluxes $\tau$ and $H$ [that is, replacing $\partial U / \partial z$ in (27) successively by $\tau$ and $H$] results in relationships for the non-dimensional momentum flux $\tau / U_{N\varepsilon}^2$,

$$\frac{\tau N}{\varepsilon} = \psi_m(\xi) \tag{29}$$

and for the non-dimensional temperature flux $H / (U_{N\varepsilon} \theta_{N\varepsilon})$,

$$\frac{\beta H}{\varepsilon} = \psi_h(\xi) . \tag{30}$$

Dimensional analysis shows that non-dimensional relationships for the standard deviation of wind speed components $\sigma_\alpha / U_{N\varepsilon}$ and air temperature $\sigma_t / \theta_{N\varepsilon}$ can be written as

$$\frac{\sigma_\alpha}{\sqrt{\varepsilon/N}} = \psi_\alpha(\xi), \qquad \frac{\sigma_t \beta}{\sqrt{\varepsilon N}} = \psi_t(\xi) , \qquad (31)$$

where $\alpha = u, v,$ and $w$. The non-dimensional relationships for the turbulent viscosity and the turbulent thermal diffusivity are:

$$\frac{K_m N^2}{\varepsilon} = \psi_{Km}(\xi), \qquad \frac{K_h N^2}{\varepsilon} = \psi_{Kh}(\xi) . \qquad (32)$$

Based on (28)-(30) and the definitions of $K_m$ and $K_h$, one can show that $\psi_{Km} = \psi_m \psi_R^{1/2} \equiv \psi_m Ri^{1/2}$ and $\psi_{Kh} = \psi_h$ in the case of the dry air. In the general case $\psi_{Kh} = \psi_h \left(1 + \frac{m}{Bo}\right)$ where $Bo$ is the Bowen ratio (the ratio of the turbulent fluxes of sensible and latent heat) and $m = 0.61 c_p / L_e \approx 0.075$ ($c_p$ is the heat capacity of air at constant pressure, and $L_e$ is the latent heat of evaporation of water).

The asymptotic behaviour of the universal functions (28)-(32) can be predicted for neutral conditions ($\xi \to 0$) and in the very stable case ($\xi \gg 1$). In the neutral case, various quantities become independent of the buoyancy parameter $\beta$ (recall that $\beta$ is included in $N$); that is, $\beta$ is no longer a primary scaling variable. This requires that $\beta$ cancels in Equations (28)-(32) in the limit $\xi \to 0$; therefore,

$$\psi_R = a_R \xi^{4/3}, \quad \psi_m = a_m \xi^{2/3}, \quad \psi_{Km} = a_{Km} \xi^{4/3}, \quad \psi_{Kh} = a_{Kh} \xi^{4/3}, \quad \psi_\alpha = a_\alpha \xi^{1/3}, \quad \psi_t = a_t \xi . \qquad (33)$$

In the very stable case, various dimensional variables become independent of $z$ ($z$-less stratification) and the universal functions (28)-(32) asymptotically approach constant values when $\xi \gg 1$:

$$\psi_R = b_R, \quad \psi_m = b_m, \quad \psi_{Km} = b_{Km}, \quad \psi_{Kh} = b_{Kh}, \quad \psi_\alpha = b_\alpha, \quad \psi_t = b_t. \qquad (34)$$

The numerical coefficients in (33) and (34) will be specified in the next section. Similarly to (19) interpolation forms can be proposed to blend between the neutral (33) and the z-less (34) asymptotic limits.

*4.2. Relationships between the universal functions for the flux-based (MOST) and N-ε based scale systems*

Although the derivation of the relationships in the Section 4.1 is independent and self-consistent, the universal functions (28)-(32) for $N$-$\varepsilon$ scaling can be expressed through the traditional MOST functions defined in Section 3 and vice versa. First, the Dougherty-Ozmidov length scale $L_{N\varepsilon} = \sqrt{\varepsilon/N^3}$ is a universal function of the Obukhov length scale $L = -\tau^{3/2}/(\kappa \beta H)$ (or, equivalently, $\xi = z/L_{N\varepsilon}$ is a universal function of $\zeta = z/L$). Substituting $\partial \theta / \partial z$ from (12) and $\varepsilon$ from (14) into $\xi = z/L_{N\varepsilon}$ yields

$$\xi = \frac{(\zeta \varphi_h)^{3/4}}{\kappa \varphi_\varepsilon^{1/2}}, \qquad (35)$$

where $\zeta$ is defined by (8). In the neutral limit ($\zeta, \xi \to 0$) Eq. (35), where $\varphi_h$ and $\varphi_\varepsilon$ are specified by (19), reduces to $\xi = (Pr_{t0}\zeta)^{3/4}/\kappa$ (i.e. $\beta$ cancels). In the very stable case ($\zeta, \xi \gg 1$), according to (19) and (35), $\xi = (\zeta \beta_h^{3/4})/(\kappa \beta_\varepsilon^{1/2})$ (i.e. z cancels). Thus in the z-less regime the Dougherty-Ozmidov length scale is linearly proportional to the Obukhov length.

Similarly to (35), substituting different variables from the appropriate MOST functions (Section 3) in the relationships (28)-(32) yields

$$\psi_m = \frac{\sqrt{\zeta\varphi_h}}{\varphi_\varepsilon}, \quad \psi_{Km} = \frac{\zeta\varphi_h}{\varphi_m\varphi_\varepsilon}, \quad \psi_{Kh} = \frac{\zeta}{\varphi_\varepsilon}, \quad \psi_\alpha = \varphi_\alpha\left(\frac{\zeta\varphi_h}{\varphi_\varepsilon^2}\right)^{1/4}, \quad \psi_t = \varphi_t\left(\frac{\zeta^3}{\varphi_h\varphi_\varepsilon^2}\right)^{1/4}. \qquad (36)$$

The universal functions $\psi_R \equiv Ri$ (28) which in MOST terms are $Ri = \zeta\varphi_h/\varphi_m^2$ [see (15)] and $\psi_h = \psi_{Kh}/(1+m/Bo)$ were defined earlier and, for this reason, are not listed in (36). Thus according to (36), the universal functions (28)-(32) derived from dimensional reasoning can be also deduced from traditional MOST functions. This allows recovering numerical coefficients in (33) and (34).

Combining Equations (28), (35) and (36) in the neutral limit yields:

$$a_R = \kappa^{4/3}, \quad a_m = \kappa^{2/3}, \quad a_{Km} = \kappa^{4/3}, \quad a_{Kh} = \frac{\kappa^{4/3}}{Pr_{t0}}, \quad a_\alpha = \beta_\alpha\kappa^{1/3}, \quad a_t = \frac{\beta_t\kappa}{Pr_{t0}}. \qquad (37)$$

In a similar manner, in the very stable case:

$$b_R = \frac{\beta_h}{\beta_m^2} = Ri_{cr}, \quad b_m = \frac{\beta_h^{1/2}}{\beta_\varepsilon}, \quad b_{Km} = \frac{\beta_h}{\beta_m\beta_\varepsilon}, \quad b_{Kh} = \frac{1}{\beta_\varepsilon}, \quad b_\alpha = \beta_\alpha\left(\frac{\beta_h}{\beta_\varepsilon^2}\right)^{1/4}, \quad b_t = \frac{\beta_t}{(\beta_h\beta_\varepsilon^2)^{1/4}}. \qquad (38)$$

Note that in the $z$-less limit, the vertical gradients of mean wind speed and virtual potential temperature are related as $\beta(\partial\theta_v/\partial z) = b_R(\partial U/\partial z)^2$; that is, $Ri = Ri_{cr} = \beta_h/\beta_m^2$.

Although relationships (36) derived in the framework of $N$-$\varepsilon$ scaling are combinations of the traditional Monin-Obukhov functions, they lead to a number of important (and elegant) relationships overlooked previously in MOST. Based on Equations (14) and (15) and the experimental fact that $\varphi_\varepsilon \cong \varphi_m$ [see Eq. (24) and the discussion in Section 3], the relationships (36) can be rewritten as follows. The universal function $\psi_m$, Eq. (29), is reduced to

$$\tau N/\varepsilon = \sqrt{Ri}. \qquad (39)$$

The equivalent form for the non-dimensional turbulent viscosity (32) is $\psi_{Km} = \psi_m Ri^{1/2}$ or

$$K_m N^2 / \varepsilon = Ri \ . \tag{40}$$

The non-dimensional relationships for the turbulent thermal diffusivity in (32) also can be written in a similar simple form. Substituting $\psi_{Kh}$ from (36) into the second Eq. (32) and combining with (16) and (24) yields

$$K_h N^2 / \varepsilon = Rf \ . \tag{41}$$

Obviously, relationships (39)–(41) are a direct consequence of Eq. (24). The non-dimensional relationships for the temperature flux (30) is $\psi_h = Rf /(1 + m / Bo)$. Similarly, non-dimensional relationships for the standard deviation of wind speed components $\psi_\alpha$ and the potential temperature $\psi_t$ in (36) are reduced to

$$\frac{\sigma_\alpha}{\sqrt{\varepsilon / N}} = \beta_\alpha Ri^{1/4} , \qquad \frac{\sigma_t \beta}{\sqrt{\varepsilon N}} = \beta_t Rf / Ri^{1/4} , \tag{42}$$

where $\varphi_\alpha = \beta_\alpha$ and $\varphi_t = \beta_t$. Relationship for $\psi_t$ also can be expressed as $\psi_t = \beta_t (Rf^3 / Pr_t)^{1/4} = \beta_t Ri^{3/4} / Pr_t$. Note that Eqs. (39)-(42), in contrast to Eqs. (28)-(32), do not contain $z$ and, thus, can be used also beyond the surface layer. Equation (35), which relates Dougherty-Ozmidov and Obukhov length scales, may also be simplified. Substituting $\varphi_\varepsilon = \varphi_m$ in (31) and combining with (15) and (16) yields

$$L / L_{N\varepsilon} \equiv \xi / \zeta = Ri^{3/4} /(\kappa Rf) = Pr_t /(\kappa Ri^{1/4}) = Pr_t^{3/4} /(\kappa Rf^{1/4}) \ . \tag{43}$$

The other two fundamental scales (26) are related to their Monin-Obukhov counterparts through $u_* / U_{N\varepsilon} = \psi_m^{1/2} = Ri^{1/4}$ and $\theta_* / \theta_{N\varepsilon} = Rf / Ri^{1/4}$.

Although most of the relationships (39)-(43) are extremely simple they are valid for the whole range $0 < Ri < Ri_{cr}$ and $0 < Rf < Rf_{cr}$ where both critical values $Ri_{cr}$ and $Rf_{cr}$ are about 0.20-0.25.

*4.3. Analysis of the SHEBA data*

Measurements of atmospheric turbulence made during SHEBA are used to plot different universal function derived earlier in the frameworks of "$N$-$\varepsilon$ scaling". Recall that the data in all plots were quality controlled as described in Section 2, and the restrictions (20) on the gradient and flux Richardson numbers have been imposed to filter out outliers and data points where the Richardson-Kolmogorov cascade is not observed. Theoretical dashed lines in various plots are based on $\varphi_m = \varphi_\varepsilon = 1 + 5\zeta$, $\varphi_h = 0.9 + 4.5\zeta$, and $\varphi_w = \beta_w = 1.3$.

Figure 2a shows typical values of the Dougherty-Ozmidov length scale $L_{N\varepsilon} = \sqrt{\varepsilon/N^3}$ observed in the stable atmospheric boundary layer. The length scale $L_{N\varepsilon}$ decreases with increasing stability from about 100 to in the *Ri* range shown. These values are in good agreement with previous estimates of $L_{N\varepsilon}$ by Hunt et al. (1985, Sec. 5), Stull (1988, Sec. 12.2.3), and Sorbjan and Balsley (2008, Figure 6).

The stability parameter $\xi = z/L_{N\varepsilon}$ versus the Monin-Obukhov stability parameter (8), $\zeta = z/L$, is shown in Figure 2b, where the dashed line is based on Eq. (35). Note, that the plot in Figure 2b by definition is not affected by self-correlation because $\xi$ shares no variables with $\zeta$ except a reference height *z*. The greater scatter of points in Figure 2b and several other plots in near-neutral conditions results from the relatively small sensible heat flux and unreliable temperature gradient measurements in this case. The relatively large scatter of the bin-averaged data for level 1 and partially for level 2 may be because levels 1 and 2 are located too close to the surface (i.e., within roughness or blending sublayers) and are, consequently, more affected by surface heterogeneity.

The universal function (28) $\psi_R = Ri$ versus $\xi = z/L_{N\varepsilon}$ is plotted in Figure 3a in a log-log representation. A similar plot $Rf$ versus $\xi$, is shown in Figure 3b. Dashed curves in Figure 3 are plotted based on parametric equations (15), (19), and (35) for upper panel and (16), (19), and (35) for the lower panel, where $\zeta$ is a parameter. In the limit $\xi \to 0$, both curves have slope 4/3; that is, $Ri, Rf \propto \xi^{4/3}$ [see also Eqs. (33) and (35)].

Plots of the non-dimensional momentum flux (29), $\psi_m = \tau N/\varepsilon$, and the non-dimensional turbulent viscosity (32), $\psi_{Km} = K_m N^2/\varepsilon$, are shown in Figures 4 and 5, respectively. Upper panels show plots versus $\xi$, where asymptotic behaviour is described by (33) and (37) in the limit $\xi \to 0$ and by (34) and (38) in the very stable case. Bottom panels in Figures 4 and 5 are plots of $\psi_m$ and $\psi_{Km}$ versus $Ri$ where the theoretical predictions are described by the simple equations (39) and (40), respectively.

Figure 6a shows the non-dimensional turbulent thermal diffusivity (32), $\psi_{Kh} = K_h N^2/\varepsilon$, versus $\xi$. Asymptotes of $\psi_{Kh}(\xi)$ are described by (33) and (37) and (34) and (38), and the dashed line is an interpolation curve. In contrast to the non-dimensional turbulent viscosity which is equal to the gradient Richardson number, $\psi_{Km} = Ri$, Eq. (40), theory predicts that the non-dimensional turbulent thermal diffusivity is equal to the flux Richardson number, $\psi_{Kh} = Rf$, Eq. (41). This dependence is shown in Figure 6b.

Figures 4b, 5b, and 6b show good agreement between experimental data and theoretical predictions (39)–(41), which are extremely simple and contain no additional calibration constants. As discussed earlier, these results are a consequence of the approximate local balance

between viscous dissipation and production of turbulence kinetic energy by the mean flow: $-<u'w'>(\partial U/\partial z) \approx \varepsilon$.

Note that, for practical applications, it is important to know how $\psi_{Kh}$ depends on $Ri$ rather than on $Rf$. Figure 7a plots the non-dimensional turbulent thermal diffusivity against $Ri$. According to Figure 7a, the scatter among different observation levels for $\psi_{Kh}$ is very high and, thus, $\psi_{Kh}$ has no universal behaviour if it is plotted versus $Ri$. This result is somewhat unexpected because a plot of $\psi_{Kh}$ versus $Rf$ (Figure 6b) looks "fine", and $Rf$ is highly correlated with $Ri$: $Pr_t = Ri/Rf \approx 0.9$ for $Ri$ and $Rf < 0.2$. One may suggest that this behaviour is associated with the influence of outliers on the bin-averaging ('spurious bin-averaging'), as briefly described in Grachev et al. (2008, p. 159-160) and discussed in detail by Grachev et al. (2012).

To limit the influence of outliers on the bin-averaging, we have imposed a prerequisite on the data in the form

$$0.5 < Ri/Ri_{SHEBA} < 2, \qquad (44)$$

where $Ri_{SHEBA} = \zeta \varphi_{h\,SHEBA}/\varphi^2_{m\,SHEBA}$ is based on the SHEBA profile functions, Eqs. (10) and (12), computed for each level separately. Sorbjan (2010) and Sorbjan and Grachev (2010) have also used the prerequisite (44), particularly for their analysis of the flux-profile relationships.

Figure 7b shows the same plots as in Figure 7a but the prerequisite (44) has been imposed on the individual data for all five levels and medians (x-symbols) to restrict the influence of the outliers. According to Figure 7b, applying the condition (44) to the data dramatically improved the situation; that is, the plot of $\psi_{Kh}$ versus $Ri$ is now much more consistent with the theoretical predictions (dashed line) as compared to Figure 7a. Sorbjan (2012) recently also discussed scatter among different observation levels in plots in which $Ri$ is the independent variable.

However the scatter in Figure 7a cannot be reduced by using Blackadar's expression for the mixing length instead of $\kappa z$ as proposed by Sorbjan (2012) simply because Figure 7 contains no $z$.

A relationship similar to (41) is widely used in oceanography to calculate the turbulent diffusivity for density $K_\rho$. The most common method of estimating $K_\rho$ was originally proposed by Osborn (1980) and is based on the stationary TKE equation, assuming a balance between the production of TKE, the buoyancy flux, and the dissipation of TKE, Eq. (23). According to Osborn (1980) $K_\rho N^2 / \varepsilon = \gamma$, where $\gamma = Rf /(1 - Rf)$ is the mixing efficiency. An upper bound on the mixing efficiency $\gamma$ is traditionally taken as $\gamma \approx 0.2$, which corresponds to $Rf = Rf_{cr} \approx 0.15$ (Osborn, 1980; Oakey, 1982). In reality, $\gamma$ is likely to vary with stratification. The mixing efficiency $\gamma$ and the Osborn method, in general, are further discussed by Peters et al. (1988), Weinstock (1992), Moum (1996), Smyth et al. (2001), Lozovatsky & Fernando (2002, 2013), among others.

Figure 8 shows plots of the normalized standard deviation of the vertical wind speed component, $\psi_w = \sigma_w \sqrt{N/\varepsilon}$ versus $\xi = z/L_{N\varepsilon}$ (Figure 8a) and versus $Ri$ (Figure 8b). Note that a relationship for the non-dimensional standard deviation of wind speed components in a general form $\sigma_\alpha /(L_{N\varepsilon} N) = \psi_\alpha(Ri)$ for $\alpha = u$ can be found in Ozmidov (1998, Eq. (10)), though without a specification for $\psi_\alpha(Ri)$. According to our study, $\psi_\alpha(Ri) = \beta_\alpha Ri^{1/4}$, Eq. (42), and $\beta_w = 1.3$ for $\alpha = w$. In fact, Eq. (42) for $\sigma_u$ in implicit form is contained in Lozovatsky and Ozmidov (1979). According to Lozovatsky and Ozmidov, $\sigma_u /(lN) = 0.8\, Ri^{-1/2}$ and $\varepsilon /(l^2 N^3) = 0.6\, Ri^{-3/2}$, where $l$ is the turbulence length scale that is associated with the wavenumber of the energy-containing eddies (spectral peak) in the spectrum of the longitudinal velocity component and should be determined experimentally. Combining these two equations from Lozovatsky and Ozmidov leads to $\sigma_u \sqrt{N/\varepsilon} = \beta_u Ri^{1/4}$ with a numerical coefficient approximately equal to one.

Our relationships (42) are also consistent with previous predictions for various parameters versus $Ri$ ($Ri < Ri_{cr}$) derived in a different context through completely different theoretical means by Rohr et al. (1988), Luketina and Imberger (1989), Weinstock (1992), Schumann and Gerz (1995), and Baumert and Peters (2000). In particular, Schumann and Gerz (1995, Figure 12) predict $eN/\varepsilon \propto Ri^{1/2}$ for $Ri < Ri_{cr}$, where $e$ is TKE. This result can be derived from Eq. (42) for $\sigma_\alpha$. The model by Baumert and Peters (2000) predicts the ratio between the Thorpe length scale (or Ellison length scale) and the Dougherty-Ozmidov length scale, $L_T/L_{N\varepsilon} \propto Ri^{3/4}$ (see also Weinstock, 1992), and the ratio of the Thorpe length scale and the buoyancy length scale, $L_T/L_B \propto Ri^{1/2}$, where $L_B = \sqrt{e}/N$ and $Ri < Ri_{cr} = 0.25$. Obviously, $L_B/L_{N\varepsilon} \propto Ri^{1/4}$ is again consistent with Eq. (42) for $\sigma_\alpha$.

## 5.   Final remarks and discussion

We develop a local similarity theory for the stably stratified boundary layer that is based on the Brunt-Väisälä frequency $N$, the dissipation rate of turbulent kinetic energy $\varepsilon$, and the buoyancy parameter $\beta$. These three variables are the governing (scaling) parameters, Eq. (25), similar to the turbulent fluxes and $\beta$, Eq. (4), used in traditional Monin-Obukhov similarity theory (MOST). The scaling parameters (25) uniquely define a system of three fundamental scales for the length, velocity, and temperature (26). A buoyancy length scale constructed from $N$ and $\varepsilon$, $L_{N\varepsilon} = \sqrt{\varepsilon/N^3}$, was originally suggested by Dougherty (1961) and independently by Ozmidov (1965) and, here, it is referred to as the Dougherty-Ozmidov length scale (in oceanography, this length scale is known as the Ozmidov length).

Based on dimensional analysis (Pi theorem) and repeating the Monin-Obukhov formalism described in Section 3, but using $N$ and $\varepsilon$ instead of the turbulent fluxes, we show that any statistics of the small-scale turbulence properly scaled with (26) are universal functions of a stability parameter defined as the ratio of height $z$ and the Dougherty-Ozmidov length scale (Section 4.1). The Dougherty-Ozmidov length scale $L_{N\varepsilon}$ is uniquely related to the Obukhov length, Eq. (35), and in the limit of $z$-less stratification, they are linearly proportional to each other. The applicability of the approach as well as MOST in stable conditions is limited by the inequalities (20), $Ri < Ri_{cr}$ and $Rf < Rf_{cr}$, where both critical values $Ri_{cr}$ and $Rf_{cr}$ are about 0.20-0.25 (cf. Grachev et al., 2013).

Because the scaling system $N$, $\varepsilon$, and $\beta$ is traditionally used in oceanography, our approach can be considered as a description of the atmospheric turbulence in "oceanographic language" or as a link between descriptions of atmospheric turbulence and oceanic vertical mixing. Equations (28)-(32), in which $\xi = z/L_{N\varepsilon}$ is the independent variable, can be used to study near-bottom oceanic turbulence (e.g., Peters and Johns, 2006; Lozovatsky et al., 2008, 2010, and references therein) or the oceanic boundary layer under pack ice (McPhee, 2008). One may assume that $N$-$\varepsilon$ scaling (Section 4) is more suitable for describing the dimensionless oceanic spectra (e.g. Lien and Sanford, 2004; Walter et al., 2011) than traditional MOST.

Our approach leads to a number of important (and simple) relationships (39)-(42) (in which $Ri$ or $Rf$ are the independent variable) overlooked previously in MOST or in oceanography [e.g. Eq. (39)]. Note that Eqs. (39)-(41), in contrast to the traditional MOST relationships, e.g. (10) and (12), have explicit forms and do not contain calibration coefficients. Moreover, Eqs. (28)-(32) contain no $z$ and, thus, can be used also far from the surface. This is due to the fact that the relationships (39)-(41) are a consequence of the approximate local balance between production of turbulence by the mean flow and viscous dissipation, Eq. (24).

Although the proposed approach is formally equivalent to the MOST (see Section 4.2), it can be used as its replacement in the case when the turbulent fluxes (primary governing variables in the MOST) are not available or cannot be measured directly. Examples of such situations include already mentioned small-scale oceanic turbulence or measurements of atmospheric turbulence by a hot-wire anemometer from aircraft, helicopters, or balloons (e.g. Muschinski et al., 2001; Sorbjan and Balsley, 2008). Thus the practical importance of the current study is associated with the description of the various small-scale turbulent statistics (including the fluxes) based on measured values of $N$ and $\varepsilon$.

It is apparent that the MOST functions $\varphi_m$, $\varphi_h$, $\varphi_\alpha$ etc. are also universal functions of the stability parameter $\xi = z / L_{N\varepsilon}$; and, vice versa, "new" functions $\psi_m$, $\psi_h$, $\psi_\alpha$, etc. are universal functions of the MOST stability parameter (8), $\zeta = z / L$. This is because of the Dougherty-Ozmidov length scale $L_{N\varepsilon}$ is uniquely related to the Obukhov length according to Eq. (35). Such a "hybrid" representation allows plotting functions which by definition are not affected by self-correlation (cf. Grachev et al. 2013, Fig 16). For example, plot of the MOST non-dimensional vertical gradient of mean wind speed (10), $\varphi_m$, versus the Dougherty-Ozmidov stability parameter $\xi = z / L_{N\varepsilon}$ is not affected by the self-correlation because $\varphi_m$ shares no variables with $\xi$ except a reference height $z$ (the plot is not shown here). At the same time, straightforward plots of $\varphi_m$ versus $\zeta$ (Figure 1a) and $\psi_m$ versus $\xi$ (Figure 4a) are affected by self-correlation.

Local similarity theory based on the scaling system $\{N, \varepsilon, \beta\}$ derived here is not the only option for a reformulation of MOST. There are many other choices to build up a local similarity theory in the SBL based on different combinations of the scaling parameters. Sorbjan (2006, 2008, 2010) formulated an approach. The choice of a certain scaling system should be based on the convenience and on the possibility of measuring specific parameters. In addition to the scaling systems by Smeets et al. (2000) and Sorbjan (2006, 2008, 2010) mentioned in Section 1, the following promising scaling systems are worth mentioning: $\{\sigma_w, \sigma_t, \beta\}$ and $\{\varepsilon, \chi, \beta\}$, where $\chi$ is the mean thermal dissipation rate. These scaling systems are associated with the buoyancy length scales $L_{wt} = \sigma_w^2/(\beta\sigma_t)$ and $L_{\varepsilon\chi} = \varepsilon^{5/4}/(\beta^{3/2}\chi^{3/4})$ (Bolgiano-Obukhov length), respectively. Note that the last scaling is equivalent to $\{C_U^2, C_T^2, \beta\}$, where $C_U^2 = 4\alpha\varepsilon^{2/3}$ and $C_T^2 = 4\alpha_T\chi\varepsilon^{-1/3}$ are the structure parameters ($\alpha \approx$ 0.5-0.6 is the Kolmogorov constant and $\alpha_T \approx$ 0.8) (e.g., Kaimal and Finnigan, 1994).

**Acknowledgements**

The U.S. National Science Foundation's Office of Polar Programs supported our original SHEBA research. During the current work NSF also supported AAG and POGP with award ARC 11-07428 and ELA with award ARC 10-19322. AAG also was supported by the U.S. Civilian Research & Development Foundation (CRDF) with award RUG1-2976-ST-10.

**Figure Captions**

Figure 1. The bin-averaged non-dimensional universal functions (*a*) $\varphi_m$, (*b*) $\varphi_h$, and (*c*) $\varphi_\varepsilon$ for five levels of the main SHEBA tower during the 11 months of measurements plotted versus the Monin-Obukhov stability parameter for local scaling $\zeta = z/L$. Both prerequisites (20) with $Ri_{cr} = Rf_{cr} = 0.2$ have been imposed on the data. The dashed lines are based on $\beta_m = \beta_\varepsilon = 5.0$, $\beta_h = 4.5$, and $Pr_{t0} = \beta_h/\beta_m = 0.9$. Individual 1-hr averaged SHEBA data based on the median fluxes for the five levels are shown as the background yellow x-symbols. SHEBA data with a temperature difference between the air (at median level) and the snow surface less than 0.5°C have been omitted to avoid the large uncertainty in determining the sensible heat flux. To avoid a possible flux loss caused by inadequate frequency response and sensor separations, a prerequisite that $U > 1$ m s$^{-1}$ has also been imposed.

Figure 2. (*a*) Behavior of the Dougherty-Ozmidov length scale $L_{N\varepsilon} = \sqrt{\varepsilon/N^3}$ (bin medians) observed in the stable atmospheric boundary layer for SHEBA data plotted against the gradient Richardson number (15). (*b*) Plot of the bin-averaged stability parameter $\xi = z/L_{N\varepsilon}$ versus the Monin-Obukhov stability parameter (8), $\zeta = z/L$. The dashed line is based on Eqs. (35) and (19), where $\beta_m = \beta_\varepsilon = 5.0$, $\beta_h = 4.5$, and $Pr_{t0} = \beta_h/\beta_m = 0.9$. Symbols and notations are the same as in Figure 1.

Figure 3. Plots of the bin-averaged (*a*) gradient Richardson number, *Ri*, and (*b*) flux Richardson number, *Rf*, (bin medians) versus the Dougherty-Ozmidov stability parameter $\xi = z/L_{N\varepsilon}$. Dashed curves are based on parametric equations (15), (19), and (35) for upper panel and on (16), (19), and (35) for the lower panel where $\zeta$ is a parameter ($\beta_m = \beta_\varepsilon = 5.0$, $\beta_h = 4.5$, and $Pr_{t0} = \beta_h/\beta_m = 0.9$). Symbols and notations are the same as in Figure 1.

Figure 4. Plots of the bin-averaged non-dimensional momentum flux (29), $\psi_m = \tau N / \varepsilon$, versus (*a*) the Dougherty-Ozmidov stability parameter, $\xi = z / L_{N\varepsilon}$, and (*b*) the gradient Richardson number, *Ri*, see Eq. (39). Symbols and notations are the same as in Figure 1.

Figure 5. Same as Figure 4 but for the non-dimensional turbulent viscosity (32), $\psi_{Km} = K_m N^2 / \varepsilon$.

Figure 6. Plots of the bin-averaged non-dimensional turbulent thermal diffusivity (32), $\psi_{Kh} = K_h N^2 / \varepsilon$, versus (*a*) the Dougherty-Ozmidov stability parameter, $\xi = z / L_{N\varepsilon}$, and (*b*) the flux Richardson number, *Rf*, see Eq. (40). Symbols and notations are the same as in Figure 1.

Figure 7. Plots of the bin-averaged non-dimensional turbulent thermal diffusivity (32), $\psi_{Kh} = K_h N^2 / \varepsilon$, versus the gradient Richardson number, *Ri*. (*a*) No restriction on *Ri*-outliers are applied; (*b*) the prerequisite (Eq. 44) has been imposed on the individual data for all five levels and medians (x-symbols) to restrict the influence of the outliers.

Figure 8. Plots of the bin-averaged non-dimensional standard deviation of the vertical wind speed component, $\psi_w = \sigma_w \sqrt{N / \varepsilon}$, versus (*a*) the Dougherty-Ozmidov stability parameter, $\xi = z / L_{N\varepsilon}$, and (*b*) the gradient Richardson number, *Ri*. Symbols and notations are the same as in Figure 1.

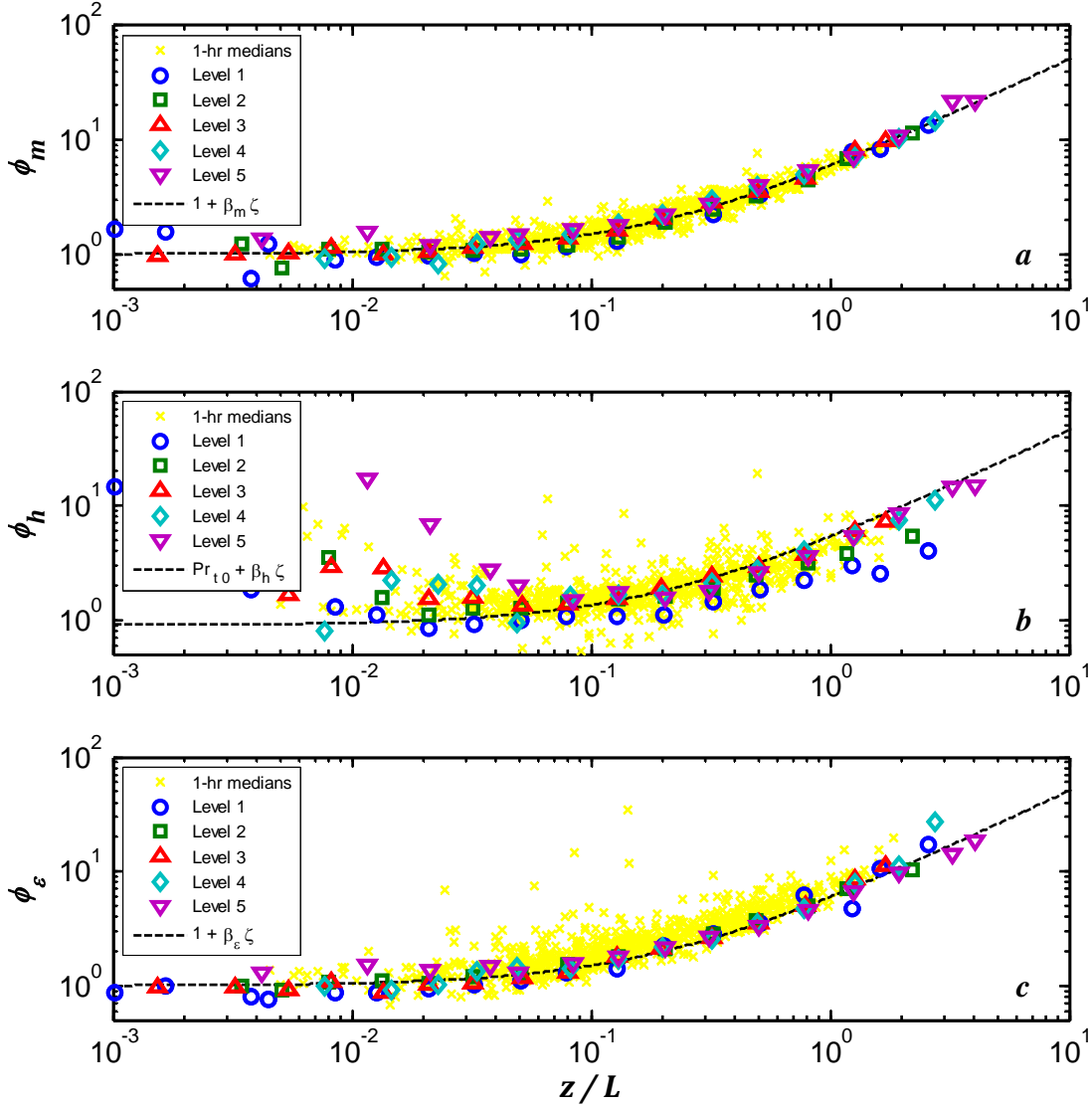

Figure 1. The bin-averaged non-dimensional universal functions (*a*) $\varphi_m$, (*b*) $\varphi_h$, and (*c*) $\varphi_\varepsilon$ for five levels of the main SHEBA tower during the 11 months of measurements plotted versus the Monin-Obukhov stability parameter for local scaling $\zeta = z/L$. Both prerequisites (20) with $Ri_{cr} = Rf_{cr} = 0.2$ have been imposed on the data. The dashed lines are based on $\beta_m = \beta_\varepsilon = 5.0$, $\beta_h = 4.5$, and $Pr_{t0} = \beta_h/\beta_m = 0.9$. Individual 1-hr averaged SHEBA data based on the median fluxes for the five levels are shown as the background yellow x-symbols. SHEBA data with a temperature difference between the air (at median level) and the snow surface less than 0.5°C have been omitted to avoid the large uncertainty in determining the sensible heat flux. To avoid a possible flux loss caused by inadequate frequency response and sensor separations, a prerequisite that $U > 1$ m s$^{-1}$ has also been imposed.

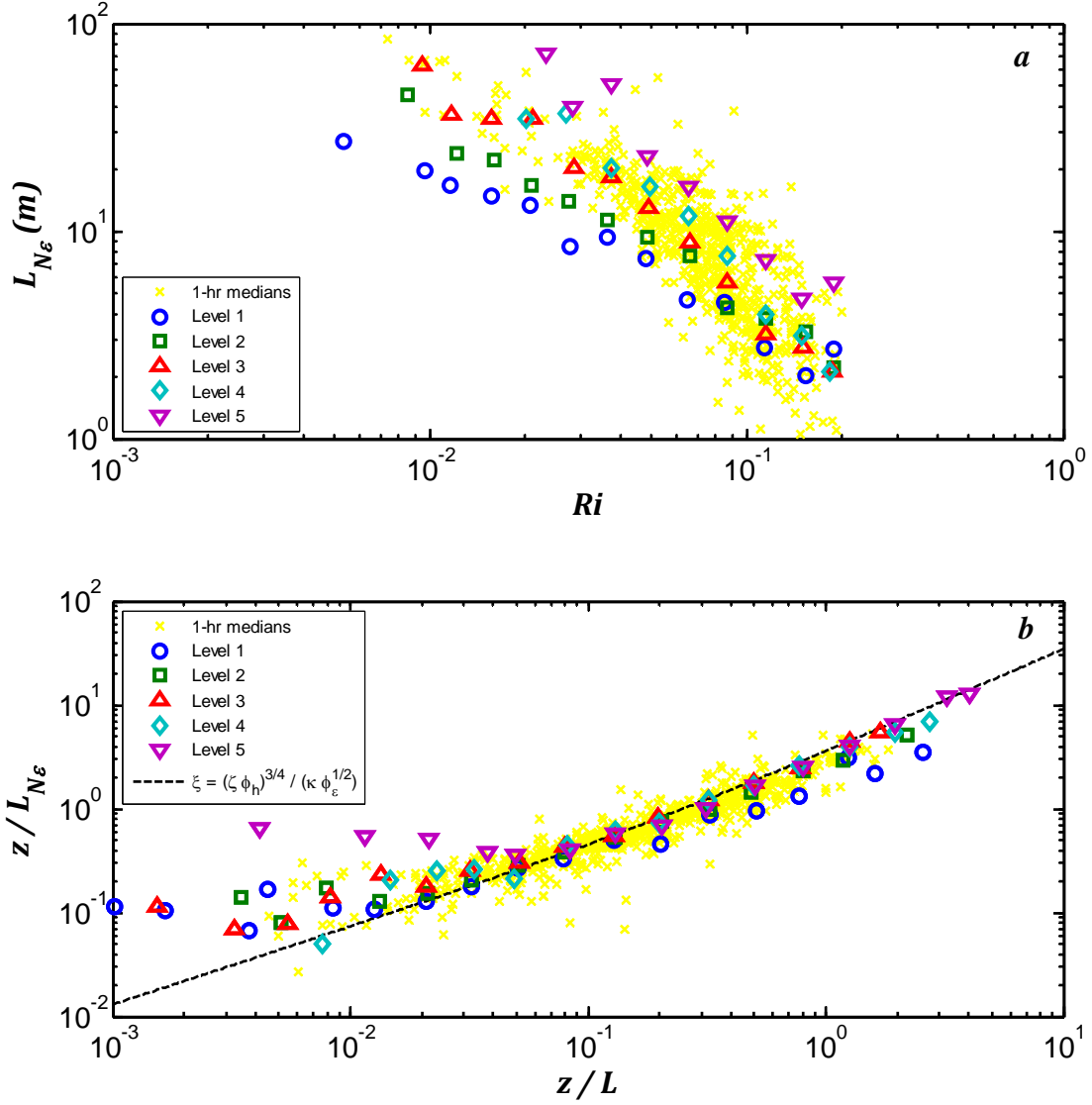

Figure 2. (*a*) Behavior of the Dougherty-Ozmidov length scale $L_{N\varepsilon} = \sqrt{\varepsilon/N^3}$ (bin medians) observed in the stable atmospheric boundary layer for SHEBA data plotted against the gradient Richardson number (15). (*b*) Plot of the bin-averaged stability parameter $\xi = z/L_{N\varepsilon}$ versus the Monin-Obukhov stability parameter (8), $\zeta = z/L$. The dashed line is based on Eqs. (35) and (19), where $\beta_m = \beta_\varepsilon = 5.0$, $\beta_h = 4.5$, and $Pr_{t0} = \beta_h/\beta_m = 0.9$. Symbols and notations are the same as in Figure 1.

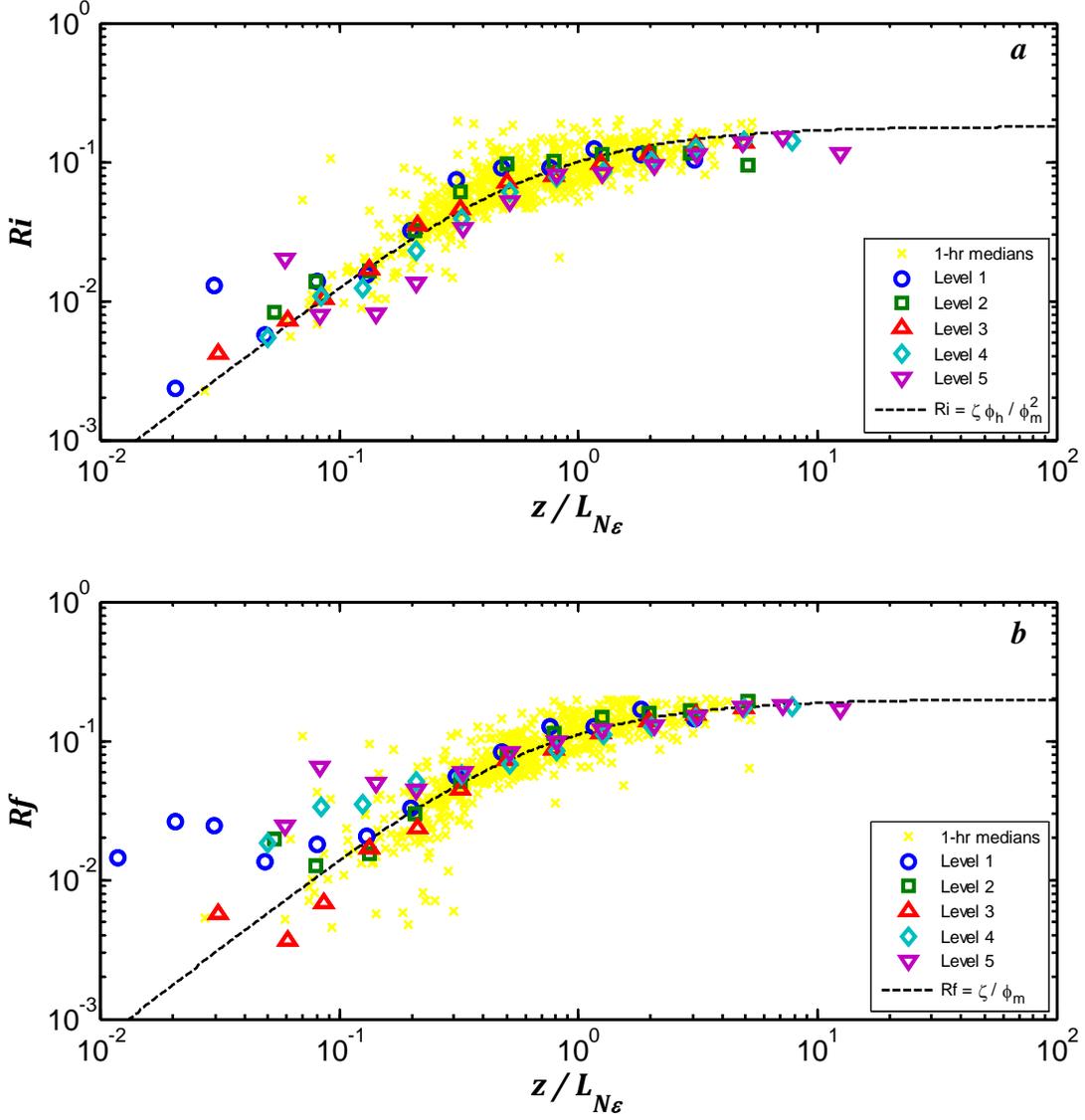

Figure 3. Plots of the bin-averaged (*a*) gradient Richardson number, *Ri*, and (*b*) flux Richardson number, *Rf*, (bin medians) versus the Dougherty-Ozmidov stability parameter $\xi = z/L_{N\varepsilon}$. Dashed curves are based on parametric equations (15), (19), and (35) for upper panel and on (16), (19), and (35) for the lower panel where $\zeta$ is a parameter ($\beta_m = \beta_\varepsilon = 5.0$, $\beta_h = 4.5$, and $Pr_{t0} = \beta_h/\beta_m = 0.9$). Symbols and notations are the same as in Figure 1.

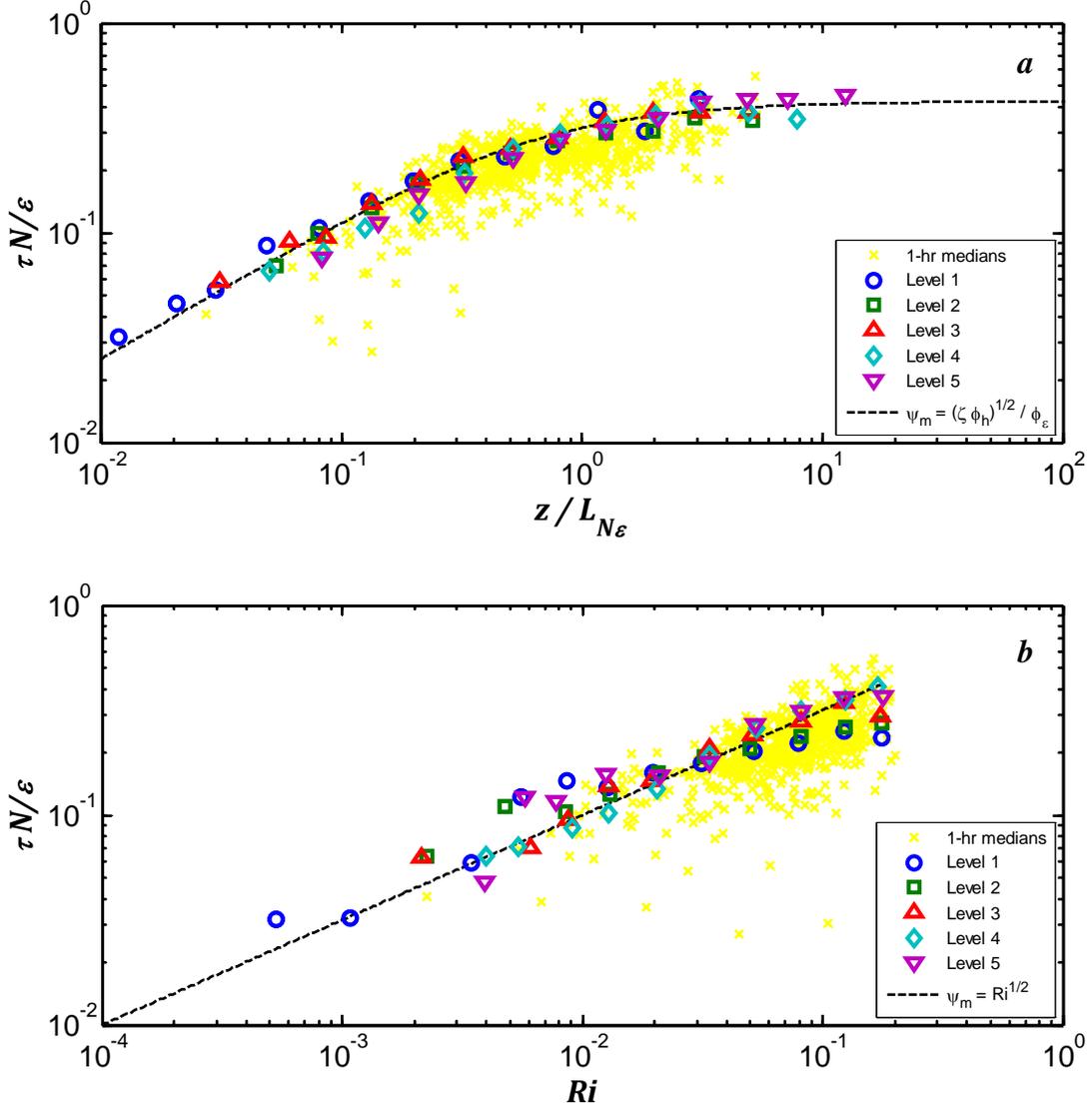

Figure 4. Plots of the bin-averaged non-dimensional momentum flux (29), $\psi_m = \tau N / \varepsilon$, versus (*a*) the Dougherty-Ozmidov stability parameter, $\xi = z / L_{N\varepsilon}$, and (*b*) the gradient Richardson number, *Ri*, see Eq. (39). Symbols and notations are the same as in Figure 1.

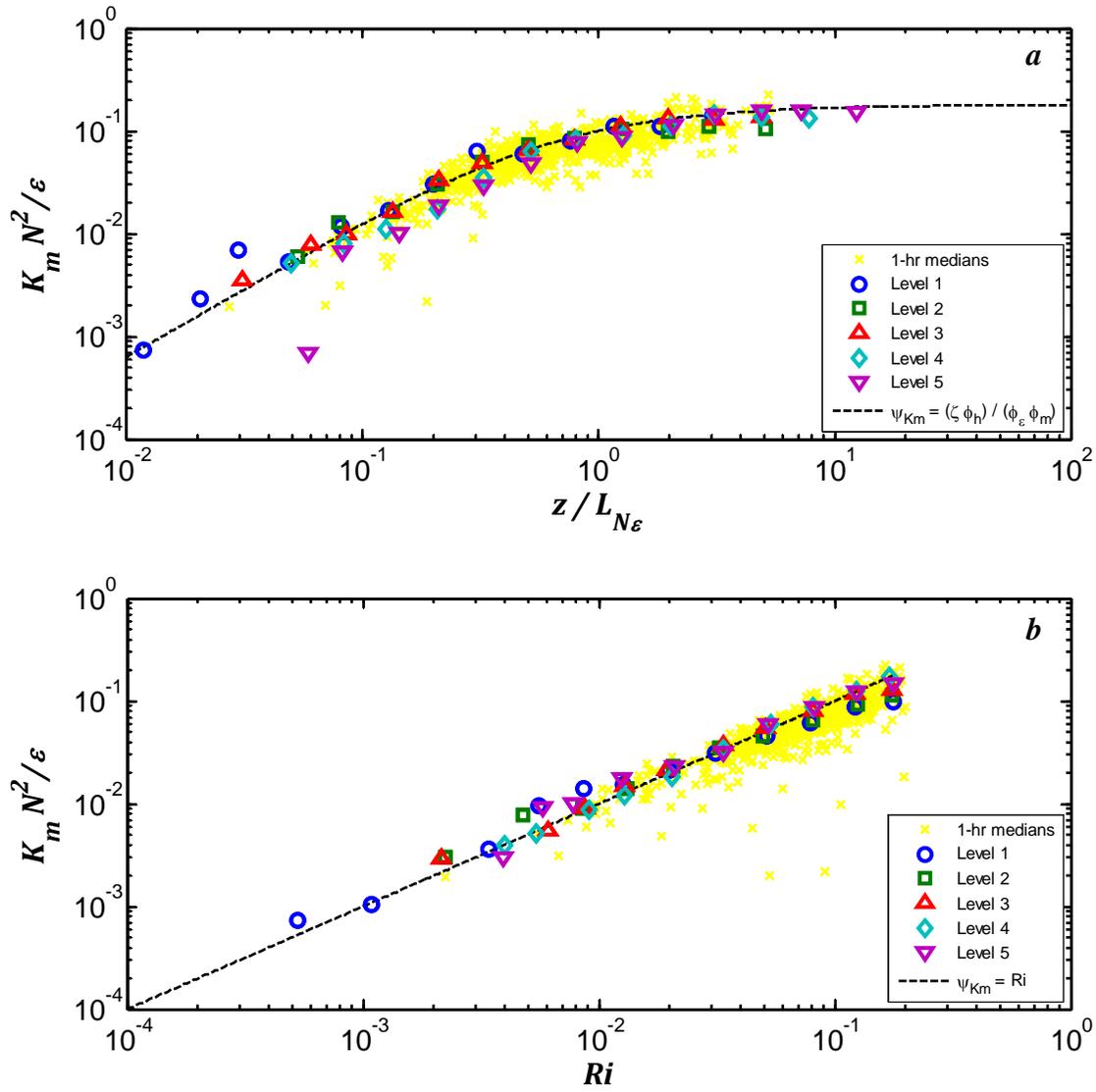

Figure 5. Same as Figure 4 but for the non-dimensional turbulent viscosity (32), $\psi_{Km} = K_m N^2 / \varepsilon$.

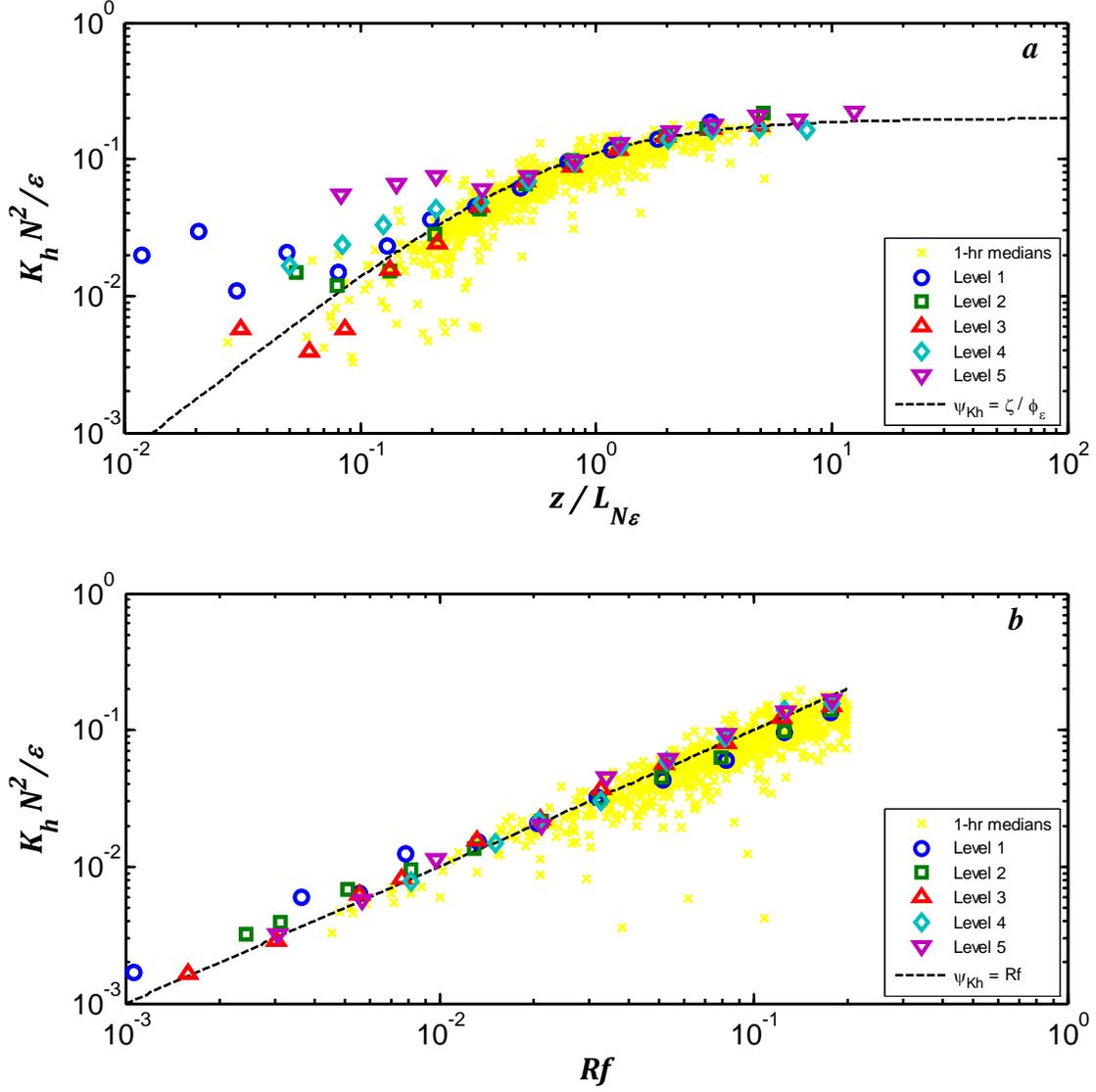

Figure 6. Plots of the bin-averaged non-dimensional turbulent thermal diffusivity (32), $\psi_{Kh} = K_h N^2 / \varepsilon$, versus (*a*) the Dougherty-Ozmidov stability parameter, $\xi = z / L_{N\varepsilon}$, and (*b*) the flux Richardson number, *Rf*, see Eq. (40). Symbols and notations are the same as in Figure 1.

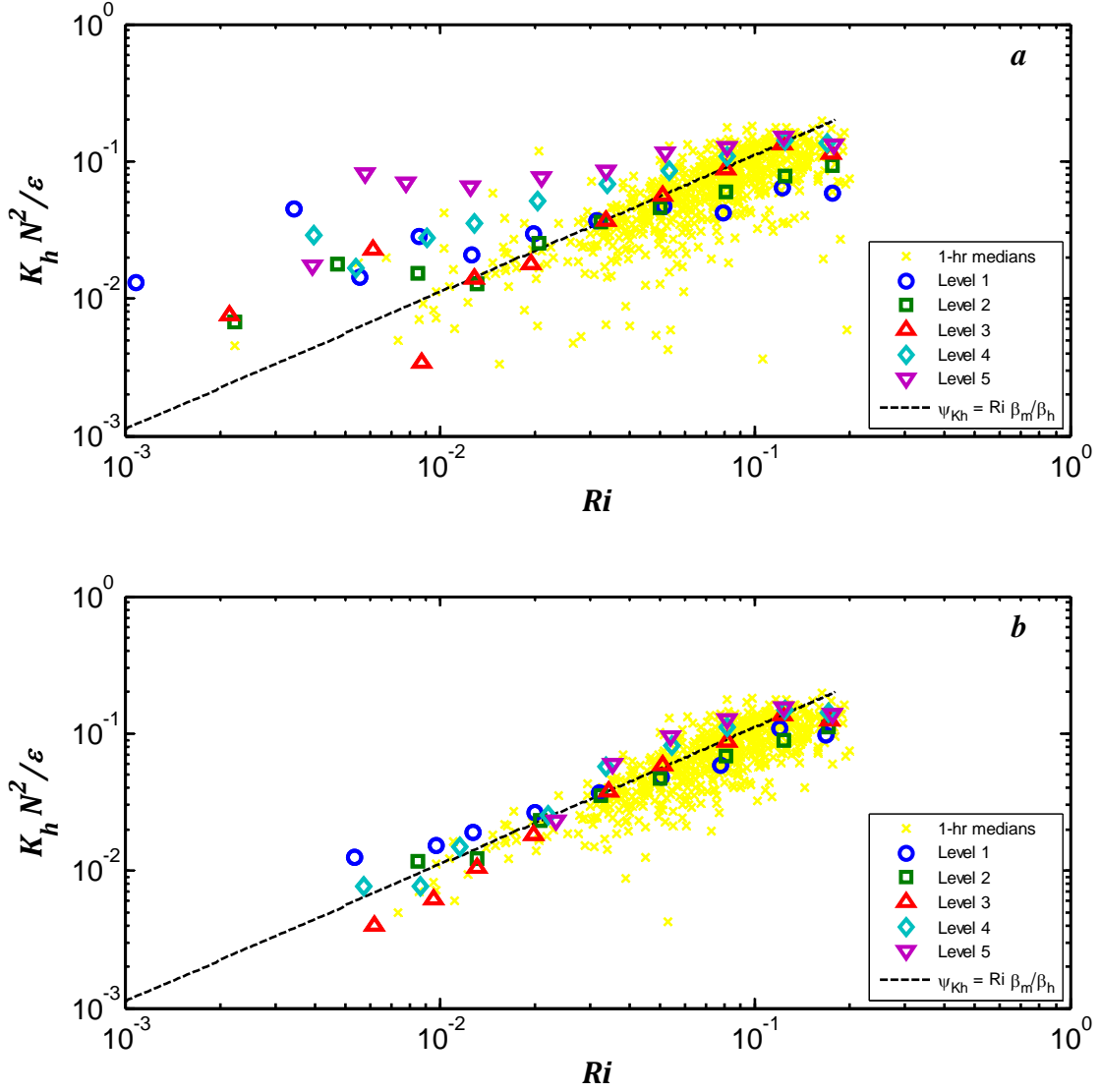

Figure 7. Plots of the bin-averaged non-dimensional turbulent thermal diffusivity (32), $\psi_{Kh} = K_h N^2 / \varepsilon$, versus the gradient Richardson number, $Ri$. (*a*) No restriction on $Ri$-outliers are applied; (*b*) the prerequisite (Eq. 44) has been imposed on the individual data for all five levels and medians (x-symbols) to restrict the influence of the outliers.

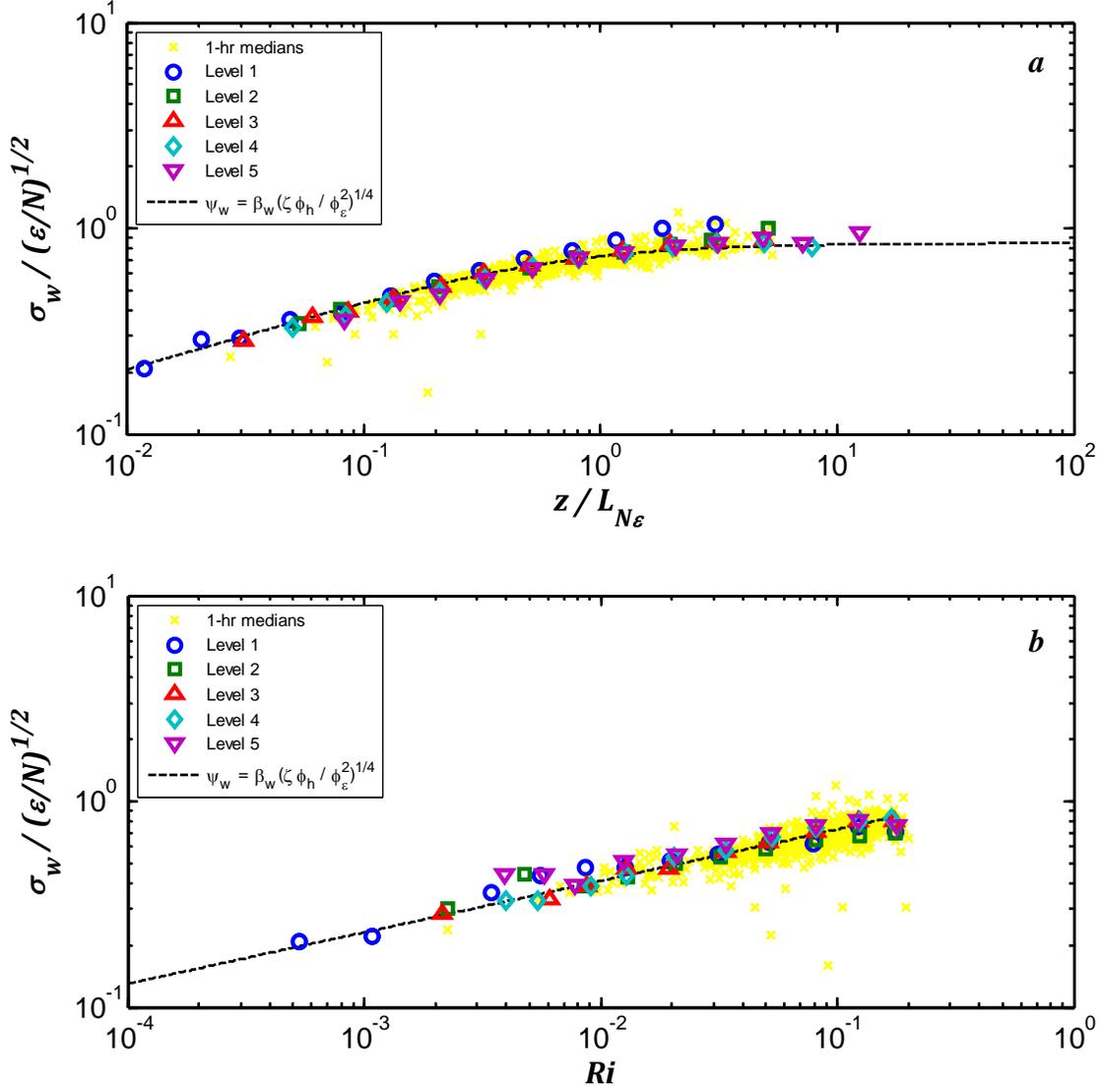

Figure 8. Plots of the bin-averaged non-dimensional standard deviation of the vertical wind speed component, $\psi_w = \sigma_w \sqrt{N/\varepsilon}$, versus (*a*) the Dougherty-Ozmidov stability parameter, $\xi = z/L_{N\varepsilon}$, and (*b*) the gradient Richardson number, *Ri*. Symbols and notations are the same as in Figure 1.